\documentclass[prl,twocolumn,showpacs]{revtex4}
\usepackage{graphicx}

\newcommand{\ordo}{{\mathcal O}}
\newcommand{\ra}{\right\rangle}
\newcommand{\la}{\left\langle}

\begin{document}
\title{Marginal scaling scenario and analytic results for a glassy compaction model}
\author{Robin Stinchcombe}
\author{Martin Depken}
\affiliation{Theoretical Physics, Oxford University, 1 Keble Road Oxford OX1 3NP, United Kingdom}
\date{\today}
\begin{abstract}
 A diffusion-deposition model for glassy dynamics in compacting granular systems is treated by time scaling and by a method that provides the exact asymptotic (long time) behavior. The results include Vogel-Fulcher dependence of rates on density, inverse logarithmic time decay of densities, exponential distribution of decay times and broadening of noise spectrum. These are all in broad agreement with experiments. The main characteristics result from a marginal rescaling in time of the control parameter (density); this is argued to be generic for glassy systems.
\end{abstract}
\maketitle
\begin{figure}[htp]
  \includegraphics[angle=0, height=0.09\textwidth, width=0.4\textwidth]{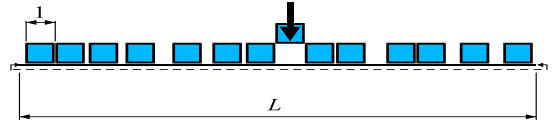}
  \caption{\label{mo}Diffusion-deposition process on a ring}
\end{figure}

Glassy dynamics occurs with similar characteristics in a remarkably diverse range of systems \cite{bok}. This letter attempts to give reasons for this similarity.\\
It begins by considering an idealized model of granular materials, which is perhaps the simplest class of real systems showing glassy behavior. The model is treated by approximate scaling and then by asymptotically exact methods. The behavior is similar to that found in simulations and in the Chicago experiments \cite{RAR,VISSGM,DRVGM,DFVGM,SRGC}\ on real granular systems.
Because we expect diverging characteristic times (while no diverging length has yet been seen in glassy systems) the scaling procedure rescales time, $t$. The consequent scaling equation for the control parameter, in this case the density $\rho$, has a particular (marginal) form. We show at the end of the letter that this leads directly to the best known characteristic features of glassy dynamics. We also argue at the end that this marginal form arises when \mbox{time scales} are excessively sensitive to changes of the control parameter, as is the case in glassy dynamics in general, so arriving at a universal scaling scenario.\\
All the following characteristics of glassy behavior \cite{bok,Angell} are shared by super cooled molecular glass-forming liquids, structural glasses, foams, colloids and even (shaken) granular systems: $(i)$ extreme slowing of rates with typically a Vogel-Fulcher dependence on control parameters (temperature, or density, etc); $(ii)$ slow decay of correlation functions and of the control parameter itself, often fitted to stretched exponential or inverse logarithmic time dependence; $(iii)$ associated aging phenomena; $(iv)$ characteristic broadening of noise spectra.\\
This has raised questions of universality, and it suggests that, despite obvious differences in fundamental mechanisms, something generic might already be present and discernable in simple models. Granular materials show all the standard characteristics when compacting under shaking, even though thermal aspects play no role~\cite{GSLG}. The simplicity of models (e.g. exclusion models) appropriate for shaken granular systems~\cite{CSGM,GRUTTP,PGG,FSDGP,TLMCDGM,CGMGCR,SEPLGA} makes them ideal candidates for such investigation, and we consider one such model here. \\
Previous investigations have shown~\cite{DFVGM,SRGC,CPADP} how a simple one-dimensional deposition-evaporation model (continuum car parking) can yield characteristic behavior. Even the noise spectrum (obtained through simulations on the model  \cite{DFVGM}) agrees well with experimental observations \cite{DRVGM,DFVGM}.\\
The model considered here is closely related to the continuum car parking model and consists of unit sized blocks performing a random walk on a ring of size $L$. They only interact through hard core potentials and as soon as a gap of unit size opens up between two adjacent blocks, an additional block will be deposited in the gap and the diffusion will continue (see Fig.~\ref{mo}).

This model allows rearrangement (diffusion)  of the grains to enable another grain to move (deposit). The model mimics the actual mechanism of rattling within and escape from/entry into rearranging cages formed by surrounding grains. Under compaction, typical free volume elements become smaller, drastically reducing the probability of finding one big enough to accommodate a block, and hence lengthening time scales. 

\begin{figure}[htp]
  \includegraphics[angle=0, width=0.35\textwidth]{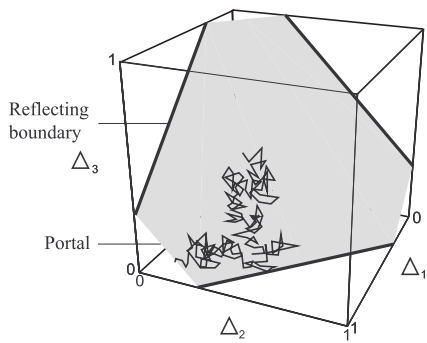}
  \caption{\label{pi}$\underline{\Delta}_3(t)$ tracing out a random walk trajectory on $\pi_3$, being reflected once at $\pi_3|_{\Delta_3=0}$ and then passing through the portal at $\pi_3|_{\Delta_1=1}$}
\end{figure}

\begin{figure}[htp]
 \includegraphics[angle=0, height=0.15\textwidth, width=0.45\textwidth]{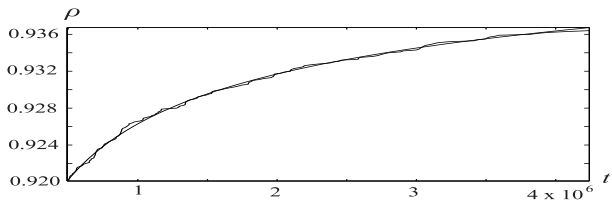}
\caption{\label{rho}Least square fit, with respect to the constant $k$ in~(\ref{rate}), of the numerical integration of~(\ref{drhodt}), compared to data from a simulation of the evolving density using parallel updates on a ring of size $L=10,000$.}
\end{figure}

\begin{figure}[htp]
 \includegraphics[angle=0, width=0.45\textwidth]{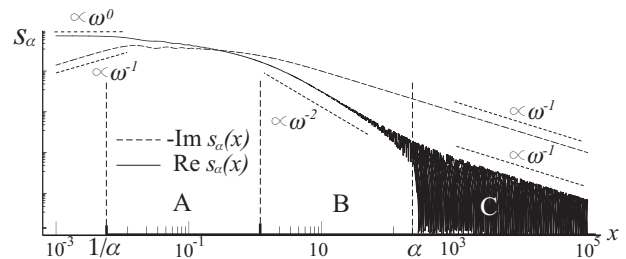}
  \caption{\label{pow}Log-log-plot of the real and imaginary parts of the power spectrum scaling function $s_{\alpha}(x)$ versus $x\equiv \omega t_A$ for $\alpha\equiv\textrm{``measuring time''}/t_A=200$. The asymptotic regions and the asymptotic behavior (see equations~(\ref{regions})) are indicated.}
\end{figure}

A qualitative scaling description of this effect is as follows. Consider elementary walkers each carrying a  gap of size $\epsilon\equiv 1-\rho$, which is the average free volume per block. For the formation of a gap of unit size, $n=1/\epsilon$ such elementary walkers have to coincide. A typical initial state has on average one elementary walker on each lattice site. Since walkers can move a distance of order $\sqrt{t}$ in time $t$, by $t\sim n^2$ walkers from $n$ sites around a particular site can have accumulated there, and do so with approximate probability per unit time
\begin{equation}
\label{probrate}
\begin{array}{c}
\left.\left(t^{-1/2}\exp(-n^2/t)\right)^n\right|_{t\sim n^2}\sim n^{-n}.
\end{array}
\end{equation}
It is straight forward to show that this result applies in any dimension. This rate of unit gap formation is equivalent to a characteristic inverse time $\tau^{-1}$ for deposition/motion from or into a cage. Equation (\ref{probrate}) implies that under rescaling of $\tau$ by a dilatation factor $b$, $n$ goes to $n'$ where $n'^{n'}=bn^n$. This equality can be written in terms of the free volume density parameter $\epsilon$ as the scaling equation
\begin{equation}
\label{scaleq}
\epsilon'=\epsilon +A\epsilon^2 \ln b+ \cdots.
\end{equation} 
In this treatment $A\sim 1/\ln \epsilon$. However, the exact discussion of the asymptotic behavior obtained below, implies that the proper form of (\ref{scaleq}) has $A$ a constant. In either form, the scaling equation has a marginal form, in which to leading order the parameter does not rescale. This marginality can also be shown to occur in the continuous car parking model and we argue below that this is a key feature of all adequate models of glassy behavior, having its physical origin in the exceptional slowing in such systems.\\
First we turn to the full analytic treatment of the model. At some instant there are $n$ blocks on the ring of size $L$. We choose to consider the gaps between adjacent blocks, $\Delta_i,\, i=1,\ldots,n$, as the dynamical variables. The vector $\underline{\Delta}_n=(\Delta_1,\ldots,\Delta_n)$ then performs a random walk on the hyper-surface (See Fig.~\ref{pi})
$$
  \pi_n=\left\{\underline{\Delta}_n\left|\begin{array}{l}\sum_{i=1}^n\Delta_i=L-n\\
 0<\Delta_i<1, \quad i=1,\ldots,n \end{array}\right. \right\}.
$$
To incorporate the hard-core interactions and the deposition process we impose the boundary conditions: $\pi_n|_{\Delta_i=0}$  reflecting (corresponding to two blocks bouncing off each other) and $\pi_n|_{\Delta_i=1}$  a {\it portal} that will transfer the system to $\pi_{n+1}|_{\Delta_i=\Delta_{i+1}=0}$ (corresponding to a gap of unit size forming and a block being deposited).\\
In what follows, unless stated otherwise, $\sim$ means ``asymptotically correct in the large system, high density limit'', and the quantity $\rho$ (or equivalently $\epsilon\equiv 1-\rho$) will be used as our control parameter. The time evolution of $\epsilon(t)$ is governed by 
\begin{equation}
\begin{array}{c}
  \dot{\epsilon}(t) \sim -\frac{1}{L\tau(t)} \label{drhodt}\quad \Rightarrow \quad \langle \dot{\epsilon}(t)\rangle=-L^{-1}\langle 1/\tau(t)\rangle.
\end{array}
\end{equation}
Here $\tau(t)$ is the time between deposition events at time $t$ (i.e. the total time spent on $\pi_{n(t)}$), for the specific realization considered. In order to estimate $\langle 1/\tau(t)\rangle$ we define the cross-section $\sigma(\epsilon)$ as the ratio between the portal area and the total boundary area (each $(n-2)$-dimensional) on each hyper-surface $\pi_n$. This will allow us to find the probability of hitting a portal, given that we hit a boundary. Through direct volume consideration one gets, after some lengthy analysis, the asymptotically exact result
\begin{equation}
\label{cross}
  \sigma(\epsilon)\sim\exp(1-1/\epsilon).
\end{equation}
The same analysis tells us that the distances between boundaries of $\pi_{L\rho}$ scale as $\epsilon$. Therefore if we let $\Delta \bar{\tau}(\epsilon)$ be the average time it takes for $\underline{\Delta}_{L\rho}$ to go between boundaries, we have $\Delta \bar{\tau}(\epsilon)\sim k \epsilon^2$, where $k$ is a constant inversely proportional to the diffusion constant of the blocks. Define, with a slight abuse of notation, $\tau(\epsilon)$ to be the time between deposition events at a given density. Since the portal cross section is very small for high densities  we expect the random walk to be essentially ergodic between the deposition events. Therefore we can get the distribution for $\tau(\epsilon)$ through considering repeated bouncing off the reflecting boundaries (blocks colliding) until finally hitting a portal (deposition event). The probability of hitting a portal on the $k$:th contact with the boundary is then $(1-\sigma)^{k-1}\sigma$, so the probability distribution for $\tau(\epsilon)$ is 
\begin{equation}
    P(\tau(\epsilon)=x)\sim(1-\sigma(\epsilon))^{x/\Delta \bar{\tau} (\epsilon)-1}\sigma(\epsilon), \label{dist2}
\end{equation}
where $x$ is a multiple of $\Delta \bar{\tau}(\epsilon)$. In the limit $\epsilon\rightarrow 0$~(\ref{dist2}) can be considered to be a continuous exponential distribution. By increasing $L$ we can, for any fixed density, have an arbitrary number of essentially independent deposition events in any finite time interval. From the law of large numbers we thus know that the $\epsilon(t)$-distribution can be made arbitrarily sharp in the large system limit. Therefore $\tau(t)\sim\tau(\la\epsilon(t)\ra)$. 
To solve~(\ref{drhodt}) for $\la\epsilon(t)\ra$ we need to calculate the average deposition rate at time $t$, $\langle 1/\tau (t) \rangle\sim\langle 1/\tau(\la\epsilon(t)\ra) \rangle$. This is easily done using~(\ref{dist2}),
\begin{equation}
\begin{array}{c}
\label{rate}
  \langle 1/\tau (t) \rangle\sim \frac{e}{k\la\epsilon(t)\ra^3}\exp(-1/\la\epsilon(t)\ra),
\end{array}
\end{equation}
which inserted into~(\ref{drhodt}) gives the asymptotic solution
\begin{equation}
\begin{array}{c}
  \label{asymp}       
   \la\epsilon(t) \ra\sim \frac{1}{\ln\left(et/kL\right)}\sim\frac{1}{\ln t}
\end{array}
\end{equation}
for large $t$. The asymptotic solution~(\ref{asymp}) is not valid at densities low enough to be suitable for comparison with practical simulations. Since the cross section~(\ref{cross}) becomes small (i.e. ergodicity holds between deposition events) already at quite moderate densities we instead compare a numerical integration of~(\ref{drhodt}) to simulation data (see Fig.~\ref{rho}). \\
To examine the power spectrum we need to include the effects of fluctuations. Since ergodicity holds on $\pi$, in the densely packed limit, we can write the Master equation for the probability distribution of $\rho$ at time $t$, $P(\rho,t)$, as,
$$
\begin{array}{c}
  \frac{\partial}{\partial t} P(\rho,t)=\frac{1}{\tau(\rho-1/L)}P(\rho-1/L,t)-\frac{1}{\tau(\rho)}P(\rho,t).
\end{array}
$$ 
This is easily solved by implementing Laplace transform techniques on the time variable. For the initial condition $\rho(0)=\rho_I$ and in the continuum limit the solution is (to the highest order of $L$ in the exponential)
\begin{equation}
\label{appro}
\begin{array}{c}
  p(\rho,t|\rho_I)\sim\tau(\rho)\sqrt{\frac{L}{2\pi T_2(\rho,\rho_I)}}\exp\left(-L\frac{(t-T_1(\rho,\rho_I))^2}{2 T_2(\rho,\rho_I)}\right)\!\!
\end{array}
\end{equation}
where $p$ is the continuum probability density and $T_n(\rho,\rho_I)=\int\limits_{\rho_I}^\rho d\rho'\tau^n(\rho')$. In order to keep the density evolution rate finite in the continuum limit $\tau$ has been rescaled as $\tau\rightarrow L \tau$.
Since $\tau(\rho)$ diverges strongly as $\rho \rightarrow 1$, the approximations used to arrive at~(\ref{appro}) break down, at any finite $L$, in the dense limit $\epsilon\lesssim 1/\sqrt{L}$. However, for large $L$ the average of $\rho$ is independent of $L$, while the variance can be made as small as we please. So for large $L$, matching of~(\ref{appro}) to a Gaussian in $\rho$ gives the correct distribution for large $L$. Using the two variable Gaussian approximation the two-time density correlation function can be calculated for a system which starts at density $\rho_I$ and is aged for a time $t_A$ from whereon measurements are performed:
\begin{equation}
\label{rr}
\begin{array}{c}
  \la \delta \rho(t_A+t)\delta\rho(t_A)\ra \sim\frac{\epsilon_I^2}{2L}\frac{1}{1+t/t_A}, \quad (t>0).
\end{array} 
\end{equation}
In the last step we have used our particular form of $\tau(\rho)$ and assumed that we are in a region where $\epsilon\sim\epsilon_A\sim\epsilon_I$, but where $\tau(\epsilon),\tau(\epsilon_A)\gg\tau(\epsilon_I)$. The above result shows that our model exhibits ``simple aging'' in the sense that the two-time correlation function is only a function of $t/t_A$. The density-density response coefficient, corresponding to perturbing the density at time $t_A$ and then measuring the response in the density at time $t_A+t$, is proportional to~(\ref{rr}) (ignoring an additive constant)~(\ref{rr}). This gives the simplest form of the fluctuation dissipation violation investigated in~\cite{CDK,LCL}. We now define the (complex) power-spectrum for the aged system as
$$
\begin{array}{c}
  S_\alpha(\omega,t_A)=\int\limits_{0}^{\alpha t_A} dt \exp(\imath\omega t)\la \delta \rho(t_A+t)\delta\rho(t_A)\ra. 
\end{array}
$$ 
Letting $E_1(z)$ denote the exponential integral, this can be written in the scaling form
$$
\begin{array}{rl}
  S_\alpha(\omega,t_A)&=t_A \frac{\epsilon_I^2}{L}s_\alpha(t_A\omega),\\
  s_\alpha(x)&=e^{\imath x}(E_1(\imath x)-E_1(\imath x(\alpha+1))).
  \end{array}
$$
Since the rates in our system are of the order of the inverse aging time (i.e. $\tau(\epsilon_A)\sim t_A$), a fairly large $\alpha$ must be considered in order to  sample the effects of aging. Therefore there are three asymptotic regimes (see Fig.~\ref{pow});
\begin{equation}
\label{regions}
  \begin{array}{lll}
   \!\!\!\!\textrm{A)}& 1/\alpha \le |x| \ll 1, & s_\alpha(x)\sim \ln(\alpha+1)-(\alpha-\ln\alpha)\imath x\!\!\!\!\!\!\!\!\!\\
    \!\!\!\!  \textrm{B)}& 1\ll |x| \ll \alpha, & s_\alpha(x)\sim \frac{1}{x^2}-\frac{\imath}{x}\\
    \!\!\!\!  \textrm{C)}& \alpha \ll |x|, & s_\alpha(x)\sim \frac{\sin(\alpha x)}{x(\alpha+1)}-\frac{\imath}{x}.
   \end{array}
\end{equation}
The reason for the lower cut-off in $x$ is that the values of $\omega<1/\alpha t_A$ are unphysical. Further we know that on short time-scales the system is time translation invariant and therefor the real part of $S_\alpha$ corresponds to the ``ordinary'' power-spectra in this limit. We naively stretch this analogy to the whole range of possible $\omega$ (and note that the the imaginary part of $S_\alpha$ stays close to the real part of $S_\alpha$ for the longer time scales (see Fig.~\ref{pow})).

Since our model only allows deposition of blocks (grains) it corresponds to the limit of very weak tapping in~\cite{DFVGM}. Taking this into account a qualitative agreement can be seen between our analytical results (see Fig.~\ref{pow}) and the experimental results presented in~\cite{DFVGM}. Further, assuming that the density increases with the depth we can interpret this as an increase of the effective aging time as measurements are done closer to the bottom of the container. Doing this we see that the bulk of the power spectra is moved towards lower frequencies while the maximum height seems to increase with aging. This is in full accord with the scaling form of $S_\alpha$.

We now discuss further the nature of time scaling in such models and the question of universality. Under rescaling of time ($t\to t'=bt$), the general renormalization group transformation~\cite{Wilson} of a control parameter $\rho$, such as the density (or temperature) can be written as $\rho \to \rho'=R_b(\rho)$. If there is a divergence of the characteristic time scale $\tau$ at $\rho=\rho^*$, then $\rho^*$ is a fixed point of the transformation. And linearizing about the fixed point using $\epsilon \equiv \rho-\rho^*$ gives in general
\begin{equation}
  \label{scaleps}
  \epsilon'=\lambda_b \epsilon+\ordo(\epsilon^2)
\end{equation}
where (from the semi-group property $R_b R_{b'}=R_{bb'}$) the eigenvalue $\lambda_b$ (if non-zero) has to have a standard power-law dependence on the time dilation factor $b$, of the form $\lambda_b=b^{1/y}$, with $y>0$. This leads to $\tau\sim|\epsilon|^{-y}$, which is ordinary critical slowing. But the essential characteristic of glassy behavior in general is that the slowing of characteristic rates with the control parameter is excessive, more extreme than with ordinary critical slowing. Then the control parameter has to change only marginally to stretch the time scales significantly. That is the case when, in (\ref{scaleps}), $\lambda_b=1$. Then higher order terms in the expansion in $\epsilon$ have to be retained, and (\ref{scaleps}) becomes of the marginal form
$$
\epsilon'=\epsilon+a\epsilon^{1+\alpha}\ln b +\ldots, \quad \alpha>1
$$
where the $b$-dependence is again set by the semi-group property~\cite{Wilson}. As was seen earlier, this form, with $\alpha=1$, applies to the diffusion-deposition model (and also the continuum car parking model, etc). Eliminating $b$ between the scaling equations for $\epsilon$ and for time gives a generalized Vogel-Fulcher form:
$$
\begin{array}{c}
 \tau\sim \exp\left(\frac{1}{\alpha a \epsilon^\alpha}\right).
\end{array}
$$
This specific equation is a direct result of the qualitative ``stronger than ordinary critical slowing'' essential characteristic of glassy dynamics. It suggests that the common Vogel-Fulcher form ($\alpha=1$) is one of various generic universal forms distinguished by the value of $\alpha$. The second commonest form would then be the exponential-inverse-squared form ($\alpha=2$), which has appeared in a glassy model with constrained dynamics~\cite{GTSDACKCSC,PS}.\\
A second characteristic of glassy systems, including the models discussed here, and responsible for meta stability, is that the control parameter is one whose mean and local fluctuations evolve with the slowed internal dynamics. Its evolution towards the fixed point value then becomes of form $\propto(\ln t)^{-1/\alpha}$ when the generalized Vogel-Fulcher form applies. This scenario applies, with $\alpha=1$, to the analytic treatment of the density evolution of the diffusion-deposition model, equation~(\ref{asymp}). The slow evolution of the control parameter is in general related to the broadening $\log$ frequency scale of the noise spectrum.  This time-scaling scenario seems equally applicable for 'thermal' glasses. It does not require a diverging length in glasses, nor a real static equilibrium transition. It does not so far explain possible/universal power laws in the noise spectrum, nor the non universal differences which distinguishes various types of ``glass''.\\
\section{Acknowledgments}
This work was supported by EPSRC under the Oxford Condensed Matter Theory grant GR/M04426. The ideas developed initially at an ITP (Santa Barbara) workshop in 1997, and RBS would like to thank the participants, and particularly the directors S. Nagel, A. Liu and S. Edwards, for sharing their insights.

\end{document}